\documentclass[reprint]{revtex4-1}
\usepackage{tikz}
\usepackage{tikz-3dplot}
\usepackage{pgfplots}
\usetikzlibrary{decorations.markings,arrows}
\pgfplotsset{compat=newest}
\usepackage{float}
\usepackage{subcaption}
\usepackage{xcolor}
\usepackage[labelfont=bf]{caption}

\usepackage{amsmath,amssymb}
\usepackage{mathtools}
\renewcommand{\vec}[1]{\mathbf{#1}}
\let\oldhat\hat
\renewcommand{\hat}[1]{\oldhat{\mathbf{#1}}}
\newcommand{\thickhat}[1]{\mathbf{\hat{\text{$#1$}}}}
\begin{document}

\title{Relativistic Aberration:\\The Transformation of a  Propagating Field}%\vspace{0.2cm} } Emanating

\author{Aran O'Hare }%\\ \small aohare20@qub.ac.uk \vspace{-0.09cm}}
\setlength{\skip\footins}{0.6cm}
\email{aohare20@qub.ac.uk}%aran.ohare@ucl.ac.uk}
% \affiliation{%
%  Authors' institution and/or address\\
%  This line break forced with \textbackslash\textbackslash
% }%

%%%%%%%\date{\today}

\begin{abstract}
The fields used to describe the influence of masses and electric charges are generally accepted to emanate at the speed of light from their sources. To obtain these fields for a moving particle which are consistent with special relativity, transformations should be applied to the emanating field in the particle's rest frame to acquire the field propagating in the particle's primed inertial frame. These transformations are of its coordinates, and also the velocity vectors of the field's propagation, which consequently lead to a change in the field's direction and as a result its density and strength.
Yet, an electric charge's field propagation velocities are neglected and not currently transformed. This omission has consequences when it comes to transforming the field from the source's rest frame to the source's primed frame, as this primed frame's field would then be inconsistent with the calculated retarded field's directions and consequently its strengths in this primed frame.
Here the retarded field of a moving point particle will be derived and from this the requirement of the velocity transforms and the aberrational effects will be shown. Hence, the full consistent inertial frame transform of the emanating field will be given. This work shows the general theoretical basis and legitimacy of the aberrational and time dilation effects on the field's density and thus strength and acts as an argument for their implementation on all luminally emanating fields.

%Here it is shown that when the velocity transform is also applied, only then can consistent transformations of a field's direction and strength be achieved between the retarded field in the source's primed frame and the field in its rest frame. 
%The results from this provide a different description of the propagating fields of forces and may bring ways of settling current experimental research into whether the field of masses or electric charges are propagated at the speed of light.
%{\color{red} here significance between the retarded field and the requirement of aberration for consistent frame transforms are shown, and provides the first fully consistent relativistic transform between the retarded field and field of particle in rest frame}

\end{abstract}

%Use showkeys class option if keyword
\maketitle

%#################################################################
\section{Introduction}
In astrophysics, when describing the observed luminosity from a relativistic plasma jet, the transformation of the radiated light's velocities, known as relativistic aberration or relativistic beaming, is used \cite{rybicki2008radiative,lind1985semidynamical}. These aberrational effects are required to explain the luminosity of the radiated light (flux of photons) from different points in the plasma, as depending on the plasma's relative velocity at a position there is a different weighting on the intensity of the number of photons received by an observer. However, coordinate transformations are not included in derivations, hence the requirement to apply the effects of aberration on the flux and a derivation of the frame transforms have not been shown.

In general, it is accepted that the electric field and its information propagate from a particle at the speed of light. However, when it comes to the relativistic transformations of electric fields, the literature applies the Lorentz transforms to the field's coordinates without transforming the field's propagation velocity \cite{EFGriffiths:1492149,EFgron2012electrodynamics,EFpurcell2013electricity,EFresnick1968introduction}. If the electric field propagates at the speed of light then relativity requires the velocity vector of this propagation at each coordinate to also be transformed. This transformation does not change the luminal speed of the propagation but does change the propagation direction. 
%Currently, the only inertial frame transform available between a moving particle’s retarded field and the field in its rest frame are the liénard-wiechert potentials, which have been used to describe synchrotron radiation \cite{wiedemann2015theory}. But these precede special relativity and do not use the relativistic retarded time required by special relativity.

Here, the retarded field of a moving particle will be derived. From this, we will show the requirement of applying both the aberrational effects (due to velocity transforms) on the field's propagation, which result in a change of the field's direction and density, and the time dilation effects on the field's density. This work shows the theoretical basis and legitimacy of both the aberrational and time dilation effects and acts as an argument for their implementation on all luminally emanating fields.
%Presented will be the first consistent relativistic transformation between the retarded field of a moving particle in an inertial frame and the field in the particle's rest frame. }
%
%the effects that emerge from the relativistic transformation of the velocities together with the coordinate transformation will be described for the system of a field propagating from a point source particle. It will be taken that in a source particle's frame of rest the field propagates at the speed of light evenly in all directions, giving a spherically symmetrical field that obeys the inverse square law, i.e. the field strength dissipates proportionately to the inverse of the square of the distance from the source particle. This field can be transformed into a primed inertial frame in which the source is moving, by using both the coordinate and velocity transformations at every point in the field. The retarded field of the moving source in the primed frame with the aberrational effects on the field strength applied is formulated. This is then shown to have equivalent field strengths and propagation directions as the previous relativistic transformation of the coordinates and velocities of the field from the source particle's rest frame.
%#################################################################
\section{The Retarded Field}
In the primed frame of a particle $q$, it is taken that the particle is moving in the positive Z-Axis only with velocity $\vec{V_q}=(0,0,v_q)$, with it currently (primed time $t'=0$) positioned at the origin. If the field emanates from the particle at luminal speed $c$. Then the field that is currently propagating through a general primed coordinate $\Vec{R'}=(x',y',z')$, is the field that was emanated from the particle when it was at a previous time and position, which can be calculated from its trajectory. These corresponding coordinates are referred to as the retarded time $T'_{ret}$ (the time when the field had emanated from the particle to reach $\Vec{R'}$ at time $t'=0$) and the particle's retarded position $\vec{P}'_{ret}=(0,0,v_q T_{ret}')$, corresponding to this retarded time. Hence the retarded field displacement that the field has propagated along is
\begin{equation}\label{displacement}
    \Vec{R'}_{ret} = \Vec{R'} - \vec{P}'_{ret} =  \begin{pmatrix}
    x'\\ y' \\ z' - v_q T_{ret}'
    \end{pmatrix}.
\end{equation}
A visual of these values are shown in figure (\ref{fig: Retarded field}) below.
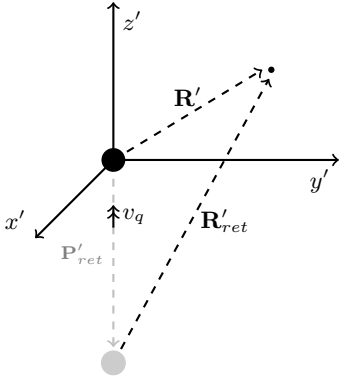
\begin{figure}
\begin{tikzpicture}[scale=3] %[,tdplot_main_coords]
%\usetikzlibrary{arrows.meta}
\draw[black, thick,->] (0,0,0) -- (1,0,0) node[anchor=north east]{$y'$};
\draw[black, thick,->] (0,0,0) -- (0,0.7,0) node[anchor=north west]{$z'$};
\draw[black, thick,->] (0,0,0) -- (0,0,0.9) node[anchor=south east] {$x'$}; 
\draw[thick,gray!50,dashed,<-] (0,-0.83,0) -- (0,0,0) node[midway,left]{\textcolor{gray}{ \scriptsize $\vec{P}'_{ret}$}};
%\node at (0,-0.65,-0.19) {$\theta_A$};
%\draw[-] (0,-0.5,0) to[bend left=90] (0.19,-0.54,0);
\draw[thick, dashed,->] (0,-0.9,0) -- (0.69,0.36,0) node[midway,black , right] {$\Vec{R}'_{ret}$};
\draw[thick, dashed,->] (0,0,0) -- (0.66,0.4,0) node[midway,above] {$\Vec{R'}$};
\fill (0,0,0) circle (1.5pt);
\fill (0.7,0.4,0) circle (0.4pt);
\draw[gray!40,fill=gray!40] (0,-0.9,0) circle (1.5pt);
\draw[black, thick,->>] (0,-0.3,0) -- (0,-0.2,0) node[midway, right]{$v_q$};
\end{tikzpicture}
\caption{Diagram of a particle's primed frame, showing a particle positioned at the origin at time $t'=0$, moving in the positive Z-direction with velocity $v_q$. The field at a general primed coordinate $\vec{R}'$, corresponds to the field that was emitted from the particle when it was at its retarded position on the Z-Axis given by $\vec{P}'_{ret} = (0,0,v_q T_{ret}')$, where $T_{ret}'$ is the corresponding retarded time. The field has propagated the retarded field displacement $\vec{R}'_{ret}$, to get to this general primed coordinate.}
\label{fig: Retarded field}
\end{figure}
The magnitude of the retarded field displacement is equal to the distance the field, propagating at the speed of light, travels in the corresponding retarded time. This gives
\begin{equation}
        \begin{split}
        \left( c T_{ret}'\right)^2 &= \|\Vec{R'}_{ret}\|^2 \\
        &= (x'^2 + y'^2 + z'^2) + v_q^2 T_{ret}'^2 - 2v_q z' T_{ret}',
%        (c^2 - v^2) T_{ret}'^2 &=  (x'^2 + y'^2 + z'^2) + 2v T_{ret}'z'
        \end{split}
\end{equation}
rearranging this, we get the quadratic
\begin{equation}
        T_{ret}'^{2} + \left(2\gamma^2\frac{v_q}{c^2} z'\right)T_{ret}' - \frac{\gamma^2}{c^2}(x'^2+y'^2+z'^2) = 0,
\end{equation}
where $\gamma = (1 - v_q^2/c^2)^{-1/2}$. Now taking the solution for the past time, and making use of the identity 
\begin{equation}\label{eq: gamma identity}
    \gamma^2 = 1+\gamma^2\frac{v_q^2}{c^2},
\end{equation}
we get the result
\begin{equation}\label{Retarded Time}
    \begin{split}
    T_{ret}' &= -\gamma^2\frac{v_q}{c^2}z' - \sqrt{\left(-\gamma^2\frac{v_q}{c^2} z'\right)^2+\frac{\gamma^2}{c^2}(x'^2+y'^2+z'^2)} \\
    &= -\gamma^2\frac{v_q}{c^2}z' - \frac{\gamma}{c}\sqrt{x'^2+y'^2+\left(1+\gamma^2\frac{v_q^2}{c^2}\right) z'^2}\\
    &= - \gamma^2\frac{v_q}{c^2}z' - \frac{\gamma}{c}\sqrt{x'^2+y'^2+\gamma^2 z'^2} \\
    &= - \gamma^2\frac{v_q}{c^2}z' - \frac{\gamma}{c}\|\vec{R}\|.
    \end{split}
\end{equation}
In the final step, we required $t'= 0$ for all coordinates, and hence used the Lorentz transform of the spatial coordinates; $\vec{R}=(x,y,z)=(x',y',\gamma z')$, to get $\|\vec{R}\|$. With this we can rewrite the Z-component of equation (\ref{displacement}) as
\begin{equation}
    z' - v_q T_{ret}' = \gamma\left(z + \frac{v_q\cdot \|\vec{R}\|}{c}\right),
\end{equation}
giving the retarded field displacement vector to be
\begin{equation} \label{retarded displacement}
    \Vec{R}'_{ret}= \begin{pmatrix}
    x\\ y \\ \gamma \left(z + \dfrac{v_q \cdot \|\vec{R}\|}{c}\right)
    \end{pmatrix} = \|\vec{R}\|\begin{pmatrix}
    \frac{x}{\|\vec{R}\|}\\ \frac{y}{\|\vec{R}\|} \\ \gamma \left( \frac{z}{\|\vec{R}\|} + \dfrac{v_q}{c} \right)\\
    \end{pmatrix}.
\end{equation}
Since the field propagates along this in the primed frame, the unit vector of the primed propagation velocity at any general primed coordinate can be worked out to be 
\begin{equation} \label{eq: unit retarded velocity}
    \thickhat{\vec{U}'} = \thickhat{\vec{R}'}_{ret} = \dfrac{1}{\text{\AA}} \begin{pmatrix}
    \frac{x}{\|\vec{R}\|}\\ \frac{y}{\|\vec{R}\|} \\ \gamma \left( \frac{z}{\|\vec{R}\|} + \dfrac{v_q}{c} \right)
    \end{pmatrix},
\end{equation}
where the factor
\begin{equation}
    \text{\AA} = \gamma\left( 1 + \frac{v_q}{c}\frac{z}{\|\vec{R}\|} \right). 
\end{equation}
Now taking the magnitude of the retarded field displacement from equation (\ref{retarded displacement}) and using equation (\ref{eq: gamma identity}), we have
\begin{equation}\label{eq: field displacement transform}
    \begin{split}
    \|\Vec{R}'_{ret}\|^2 &= x^2+y^2 + \gamma^2\left( z^2 +\frac{v_q^2}{c^2}\|\vec{R}\|^2 + 2 \frac{v_q}{c}z \|\vec{R}\| \right) \\
    &= \gamma^2 \|\vec{R}\|^2 + \frac{v_q^2}{c^2}\gamma^2z^2 + 2 \frac{v_q}{c}\gamma^2 z \|\vec{R}\| \\
    &= \gamma^2\left( \|\vec{R}\| + \frac{v_q}{c}z \right)^2\\
    &= \gamma^2\left( 1 + \frac{v_q}{c}\frac{z}{\|\vec{R}\|} \right)^2 \|\vec{R}\|^2 \\
    &= \text{\AA}^2 \|\vec{R}\|^2
    .
    \end{split}
\end{equation}
Since the luminal speed of the propagation will be same in both frames we have
\begin{equation}\label{eq: retarded field displacement transform}
\begin{split}
   c = \frac{\|\Vec{R}\|}{T_{R}} &= \frac{\|\Vec{R}'_{ret}\|}{T_{ret}'}\\
    &=  \frac{\text{\AA}\|\Vec{R}\|}{T_{ret}'} \\
    T_{ret}' &= \text{\AA} T_{R},
\end{split}
\end{equation}
where $T_{R}$ is the proper (particles rest frame) time the field takes to propagate to $\vec{R}$ from the particle. Time and length are stretched in the primed frame along $\vec{R}'_{ret}$ relative to the proper frame along $\vec{R}$ by a factor $\text{\AA}$, leading to the relative radial density of the field in the primed frame to that of the proper frame, which we will refer to as the radial field strength weighting, given as
\begin{equation}\label{eq: radial weighting}
    W_\rho = \frac{1}{\text{\AA}}.
\end{equation} 
%%%%%%%%%%%%%%%%%%%%%%%%%%%%%%%%%%%%%%%%%%%%%%%%%%%%%%%%%%%%%
% \subsection{The Field Strength}\label{ch: field strength retarded}
% %
% The field strength $f$, in the proper frame, is taken to obey Coulumb's law, i.e. it follows the inverse square law $f\propto \|\vec{R}\|^{-2}$. If this is true then to have a consistent transform between the proper and primed frame we would require the prime frame's strength to be inversely proportional to the inverse of the square of the retarded displacements magnitude with the factor $\text{\AA}$ applied, $f' \propto \text{\AA}^2\|\vec{R}_{ret}'\|^{-2} = \|\vec{R}\|^{-2} $. But so far we do not have any physical reason to apply this factor to the inverse square law. Fortunately, special relativity can explain this when the propagation velocity transformation is applied, as this turns out to be an effect due to relativistic aberration. This primed field would also be proportional to the Doppler weighting $W_D$.
%#################################################################
\section{Aberration}
\begin{figure*}
\begin{subfigure}{.49\textwidth} 
  \includegraphics[width=\textwidth]{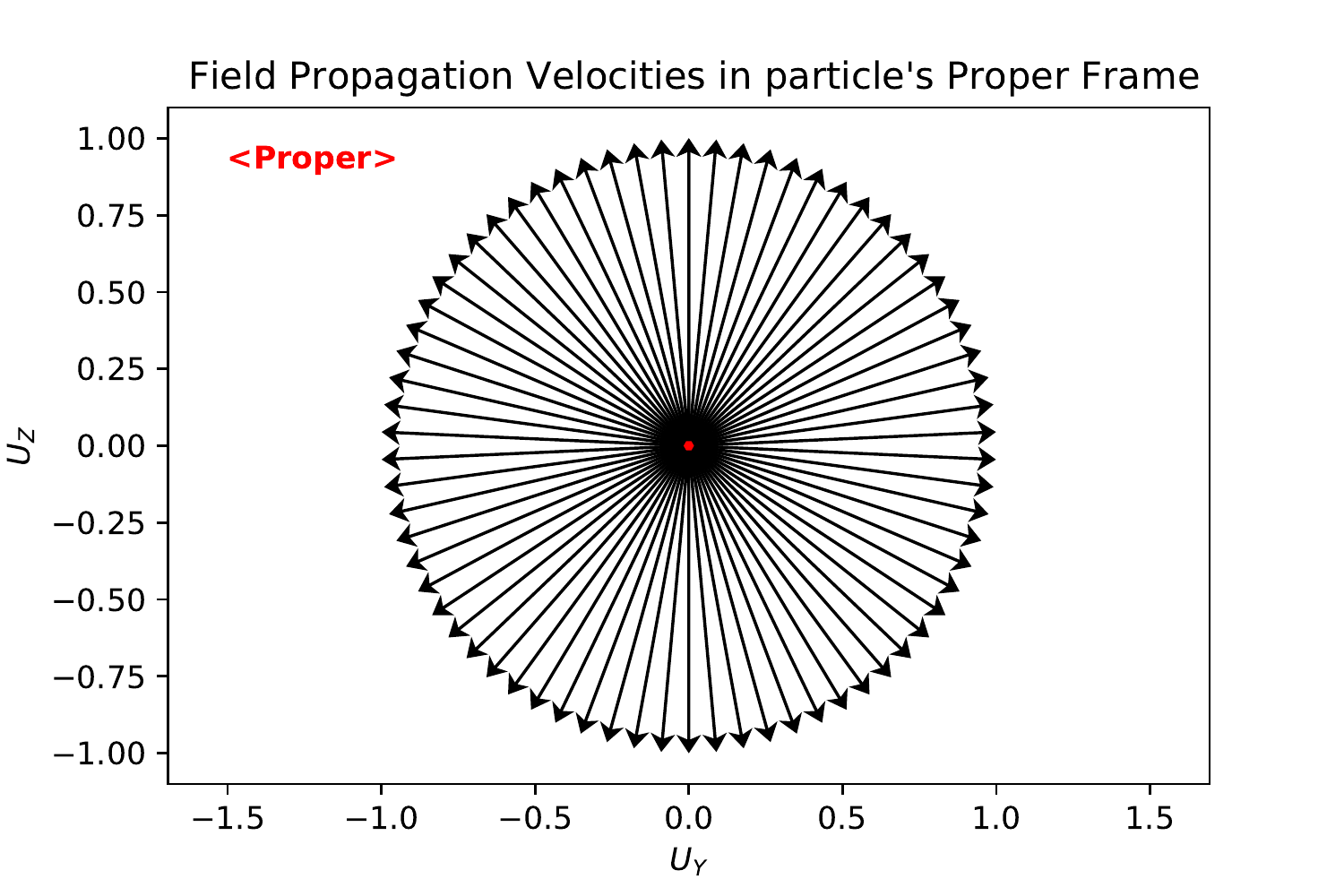}
%  \caption{Proper frame}
%  \label{fig:sub1}
\end{subfigure}
\begin{subfigure}{.49\textwidth}
  \includegraphics[width=\textwidth]{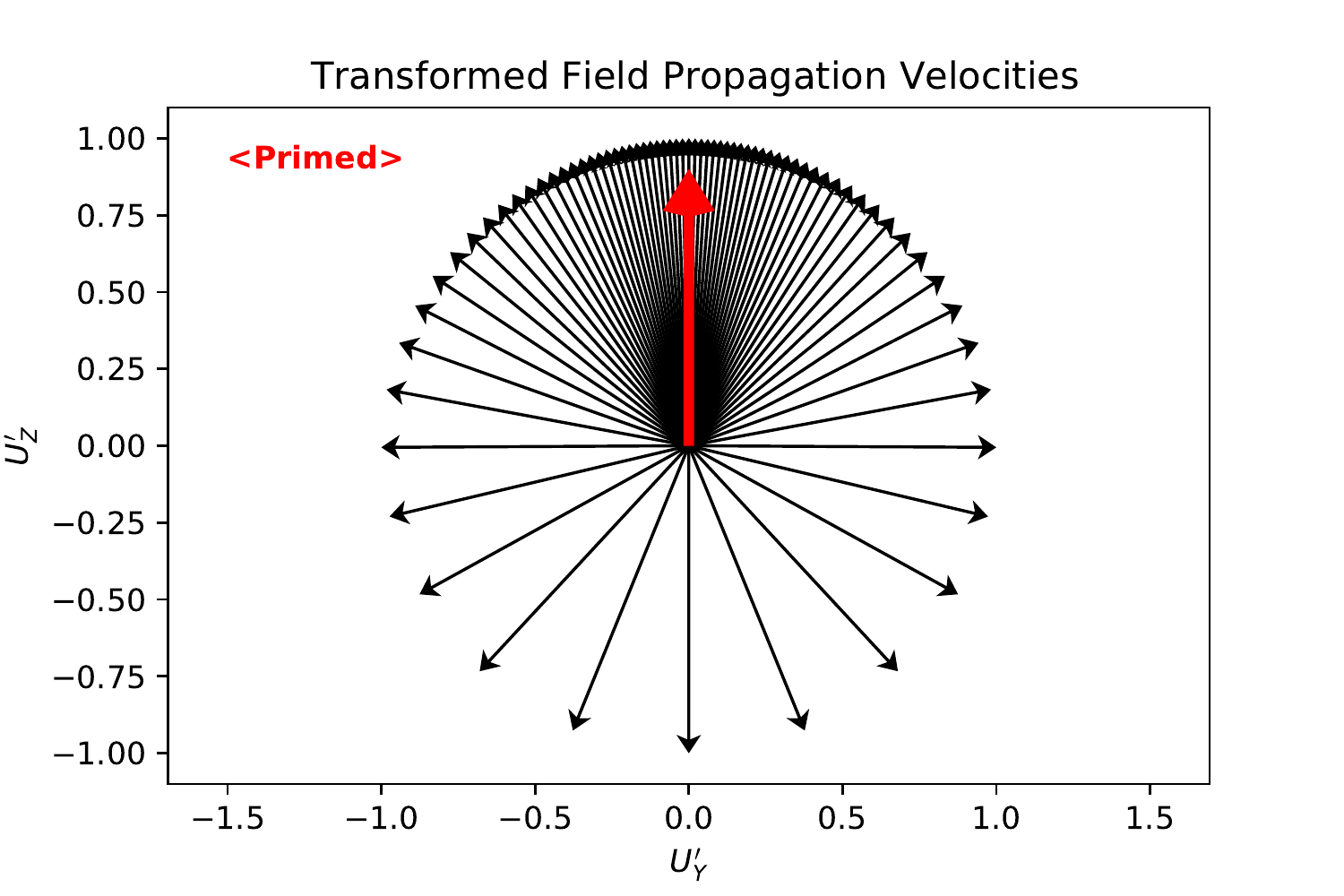}
%  \caption{Primed frame}
%  \label{fig:sub2}
\end{subfigure}
\caption{Graphs of the field propagation velocities from a particle in its proper (left) and primed (right) frames, with the same evenly distributed proper angles. The black arrows show the magnitude and direction in the yz-plane of the propagation velocity in each frame for the corresponding proper angle, and the central red arrow shows the velocity of the particle in the primed frame $v_q=0.9$.}
\label{fig: field propagtion}
\end{figure*}
The propagation of the field in the proper frame, with the particle at rest, is taken as evenly distributed with its strength spherically symmetrical, and with the magnitude of its velocity equal to the speed of light $c$, in all directions, as shown in the left graph of figure (\ref{fig: field propagtion}) below. If the field propagates,
%instead of moving rigidly with the particle at the same velocity,
then a velocity transform of this propagation at all coordinates should also be performed in order to change inertial frame. The luminal velocity of a field emanating from a particle in its own proper frame, can be written in spherical polar coordinates as
\begin{equation}
    \vec{U} = 
    \begin{pmatrix}
        U_x\\ U_y\\ U_z
    \end{pmatrix}
    = c
    \begin{pmatrix}
        \sin\theta\cos\phi\\ \sin\theta\sin\phi\\ \cos\theta
    \end{pmatrix},
\end{equation}
where $\theta \in [0,\pi] $ and $\phi \in [0,2\pi)$ are the angles from the Z-Axis and X-axis respectively. It is taken that the primed frame is moving as it did in the previous chapter, with velocity $\vec{V}=(0,0,-v_q)$ relative to the particles proper frame. Therefore we only require a transformation of the Z-component to get the primed propagation velocity. Using the equation for the relativistic transform of a general velocity vector between two inertial frames, we have
\begin{equation}\label{velocity transform}
    \begin{split}
    \vec{U}'  &= \dfrac{1}{\gamma} \dfrac{\vec{U} + \Big[\dfrac{\gamma-1}{\|\vec{V}\|^2}(\vec{U}\cdot \vec{V})- \gamma \Big] \vec{V}}{1 - \dfrac{\vec{U}\cdot\vec{V}}{c^2}}\\
    &= \dfrac{c}{\text{\AA}} \begin{pmatrix}
     \sin\theta\cos\phi\\  \sin\theta\sin\phi\\ \gamma\left(  \cos\theta + \dfrac{v_q}{c} \right)
    \end{pmatrix} = \dfrac{c}{\text{\AA}} \begin{pmatrix}
    \frac{x}{\|\vec{R}\|}\\ \frac{y}{\|\vec{R}\|} \\ \gamma \left( \frac{z}{\|\vec{R}\|} + \dfrac{v_q}{c} \right)
    \end{pmatrix},
    \end{split}
\end{equation}
with
\begin{equation}\label{eq: aberration formula in carteasian}
    \text{\AA} = \gamma\left(1+\dfrac{v_q}{c}\cos\theta\right) = \gamma\left( 1 + \frac{v_q}{c}\frac{z}{\|\vec{R}\|} \right),
\end{equation} 
where the propagation velocity is given in spherical polar and Cartesian coordinates respectively. It can be seen from equation (\ref{eq: unit retarded velocity}) that this propagation velocity's direction is the same as what is given by the retarded field. The field's propagation direction has been aberrated, while its propagation speed remains equal to $c$, which can be shown by taking the magnitude of the primed velocity vector. 
%In other words the aberration formula from equation (\ref{Cosine transform}) of the cosine of the angle $\theta$ between the Z-axis and general proper displacement $\vec{R}$ (corresponding to $\Vec{R}'$) gives the cosine of the retarded angle between the Z-axis and the retarded field displacement $\vec{R}'_{ret}$. 
%The directions of the proper field and its transformed primed field (equivalent to the retarded field) are shown in figure (\ref{fig: field Propagation Directions}) respectively. 
This aberration leads to a higher concentration of the field in the direction of the particle's motion, as shown in the right graph of figure (\ref{fig: field propagtion}) above. The propagation directions of the field as seen in figure (\ref{fig: field propagtion}), are transformed such that the angles between them and the Z-Axis, have the aberrational relationship
\begin{equation}\label{Cosine transform}
    \cos\theta' = \frac{\vec{U}'_z}{\|\vec{U}'\|} \equiv \frac{\vec{R}'_{ret_z }}{\|\vec{R}'_{ret}\|} =  \dfrac{\cos\theta + \dfrac{v_q}{c}}{1+\dfrac{v_q}{c}\cos\theta}.
\end{equation}
This is the relativistic aberration formula \cite{einstein1905electrodynamics}, it shows how the field's propagation direction transforms, it can be used to give $\text{\AA}$ in primed terms, by rearranging this for $\cos\theta$ and substituting into equation (\ref{eq: aberration formula in carteasian}), leading to
\begin{equation}\label{eq: aberration formula in primed terms}
    \text{\AA} = \frac{1}{\gamma\left(1-\dfrac{v_q}{c}\cos\theta^{'}\right)} = \frac{1}{\gamma\left( 1 - \dfrac{v_q}{c}\dfrac{\vec{R}'_{ret_z}}{\|\vec{R}'_{ret}\|} \right)}.
\end{equation} 
% The inverse transform of the cosine is given as 
% \begin{equation}\label{inverse Cosine transform}
%     \cos\theta =  \dfrac{\cos\theta' - \dfrac{v_q}{c}}{1-\dfrac{v_q}{c}\cos\theta'},
% \end{equation}
% which is achieved by the rearrangement of the previous equation.
%%%%%%%%%%%%%%%%%%%%%%%%%%%%%%%%%%%%%%%%%%%%%%%%%%%%%%%%%%%
\subsection{The Field Strength's Dependence on Angular Density}
The proper frame's differential solid angle element  
\begin{equation}
    d\Omega = \sin{\theta} d\theta d\phi,
\end{equation}
encompasses a certain amount of the field, this is the same amount of the field that is encompassed by the coinciding aberrated differential solid angle
\begin{equation}
    d\Omega' = \sin{\theta'} d\theta' d\phi'.
\end{equation}
We can calculate this element by differentiating both sides of equation (\ref{Cosine transform}) with respect to $\theta$ \cite{hogg1997special}, which gives
\begin{equation}
    \sin{\theta'} d\theta' =   \dfrac{1-\dfrac{v_q^2}{c^2}}{\left(1+\dfrac{v_q}{c}\cos{\theta}\right)^2} \sin{\theta} d\theta = \frac{1}{\text{\AA}^2} \sin{\theta}d\theta.
\end{equation}
Using this and $d\phi'=d\phi$ (as the angle $\phi$ is always perpendicular to the motion of the particle and hence unaffected by transformation) we have the solid angle in the primed frame given as
\begin{equation}
    d\Omega' = \frac{1}{\text{\AA}^2} \sin{\theta} d\theta d\phi.
\end{equation}
%The overall amount of field is conserved, as can be seen when integrating over the corresponding solid angle in either frame, both of which gives the same value of $4\pi$.
%  proportional to the amount of the field in the given proper differential element of the solid angle, and
The relative primed field strength at a given angle is taken as being proportional to the amount of the field per solid angle in the primed frame relative to that in the proper frame, referred to here as the aberrational field strength weighting, given as
\begin{equation} \label{eq: aberrational field weighting}
    W_\Omega = \frac{d\Omega}{d\Omega'} = \text{\AA}^2.
\end{equation}
%

% This aberrational effect is the physical meaning to what was needed in section (\ref{ch: field strength retarded}) to explain the transformation of the inverse square law component of the field strength between $\|\Vec{R}'_{ret}\|^{-2}$ and $\|\Vec{R}\|^{-2}$. Together the aberration and Doppler weightings allow for the full transformation between the proper and retarded field strengths.
%#################################################################
\section{The Field Transformation}
In the particle's proper frame the field strength's magnitude $f$, obeys the inverse square law, Giving
\begin{equation} \label{inverse square law}
    f =  \frac{k}{\|\vec{R}\|^2},
\end{equation}
where $k$ is a constant for the particle. In the primed frame the field strength at a coordinate will be equal to the inverse square of the retarded field displacement with constant $k$ and the field strength weightings from equations (\ref{eq: radial weighting}) and (\ref{eq: aberrational field weighting}) applied
\begin{equation}
   f' = W_\rho W_\Omega \frac{k}{\|\Vec{R}'_{ret}\|^2} = \text{\AA} \frac{k}{\|\Vec{R}'_{ret}\|^2}.
\end{equation}
It can be noted that both field strength weightings together give $\text{\AA}$, which is the same as the generalised Doppler factor. Now using this equation of the field strength and the direction of the field's propagation which is the unit vector for the retarded field displacement, we have the primed frame's field described by the vector
\begin{equation}\label{eq: retarded field}
    \vec{f}' = f'  \thickhat{\vec{R}'}_{ret}  = \text{\AA} \dfrac{k \Vec{R}'_{ret}}{\|\Vec{R}'_{ret}\|^3} .
\end{equation} 
%#################################################################
% \section{Example of the Electric Field}
% If the electric field propagates at the speed of light with its direction parallel to the direction of propagation, then the field transformations in this paper should be applied to the electric field. Consequently we can apply equation (\ref{eq: Primed Field Vector}) to the electric field, where coulombs law applies to an observer experiencing the field in the proper frame of the particle, as there are no time varying fields or potentials. Due to the lack of symmetry of the primed electric field of a particle in the Z-direction, an infinitely long current carrying wire lying along the Z-axis, will have a non zero Electric field in the Z-direction in the wire's frame. This would mean, that movement would be induced for a stationary particle in this wire's frame to move in the Z-direction. So if an electric field propagates at the speed of light and adheres to the rules of special relativity, the relativistic transform of the electric field is different from the current understanding due to the additional use of these aberrational effects. If equation (\ref{eq: Primed Field Vector}) is not correct for the electric field, then either the electric field does not propagate at the speed of light or this field propagation does not transform by the laws of special relativity.
%#################################################################
\section{Discussion}
If the electric field propagates at the speed of light and special relativity is applied, then the aberrational and time dilation effects are required, and hence the electric field will be described differently than previous relativistic transforms, described by equation (\ref{eq: retarded field}).\newline
One consequence of aberration is that if a particle approaches the speed of light, then every proper propagation direction of the field from the particle would, in the primed frame, approach the direction of the particle's motion, the retarded position of the particle would also tend to negative infinity. Given this and that the particle's motion is approaching the speed of the luminal propagation, then there would be no field outside of where the particle is located. Therefore if a photon did have a emanating field itself, it would be expected that a field would not exist outside the space that the photon occupies. \newline
%Another reflection is that if the propagation of the field was instead inwards, i.e. $c\rightarrow -c$ for equation (\ref{Cosine transform}), then this would be equivalent to the inverse transformation for the outward field propagation shown in equation (\ref{inverse Cosine transform}). We would then have that the field propagating inwards from a certain direction would have the same weighting as a field flowing out would have in the same direction. Another way of stating this is that a moving source's absorption of an inwardly propagating field at a given angle has the same weighting factor as the emitted outwardly propagating field at the same angle. 
The aberrational weighting from the transformations of a field in this paper can also be applied to the average flux of photons in all directions from a moving radiating source (instead of the field strength) and might provide another way of explaining the distribution of synchrotron radiation, where the Liénard – Wiechert field equations which include similar factors have been used to explain the phenomena \cite{wiedemann2015theory}.
%\vspace{2cm}
%Also, If a magnetic part of the field is required to be invoked, it would have the same field strength weighting as the electric field and follow the same rules for its direction relative to the electric field that is in the direction of the field's propagation.
%%%%%%%%%%%%%%%%%%%%%%%%%%%%%%%%%%%%%%%%%%%%%%%%%%%%%%%%%%%%%%%%%%%%%%
% \newpage
% \phantom{d}
% \newpage
%\section*{Acknowledgments}
\begin{acknowledgments}
The many discussions with Patrick Mullan helped pave the way to this work.
\end{acknowledgments}
\vspace{0.3cm}
%%%%%%%%%%%%%%%%%%%%%%%%%%%%%%%%%%%%%%%%%%%%%%%%%%%%%%%%%%%%%%%%%%%%%%
% \appendix
% \section{Graphs}
% \counterwithin{figure}{section}
% %
% \begin{figure}[H]
% \centering
%   \includegraphics[width=0.5\textwidth]{Field_Strength_Ratio.pdf}
% \caption{Graph showing the strength of a field in the primed frame at each primed angle relative to the corresponding proper field strength, taking into account both the aberrational and Doppler effects.}
% \label{fig: field Flux}
% \end{figure}
% %
% \begin{figure}[H]
% \begin{subfigure}{.49\textwidth}
%   \includegraphics[width=\textwidth]{Field_Directions_Proper.pdf}
% \end{subfigure}
% \begin{subfigure}{.49\textwidth}
%   \includegraphics[width=\textwidth]{Field_Directions_Aberrated.pdf}
% \end{subfigure}
% \caption{Graph showing the field's propagation direction from a source particle positioned at the origin in the proper (top) and primed (bottom) frames, with the field's propagation direction in the primed frame also being equivalent to that of the retarded field. The black arrows show the direction of the propagating field at each coordinate in the yz-plane, and the central red arrows showing the source particle's direction of motion in each frame, with $v_q=0.9$.}
% \label{fig: field Propagation Directions}
% \end{figure}
%%%%%%%%%%%%%%%%%%%%%%%%%%%%%%%%%%%%%%%%%%%%%%%%%%%%%%%%%%%%%%%%%%%%%%
% \bibliographystyle{unsrt}
% \bibliography{Bibliography.bib}

\end{document}